\documentclass[ALICE,manyauthors]{cernphprep}
\usepackage{color} 
\usepackage{bm}
\usepackage{cite}

\usepackage{rotating}
\usepackage{lineno}
\usepackage{hyperref}
\usepackage{siunitx}
\usepackage{gensymb}

\newcommand{\jpsi}{{{\rm J}/\psi}}
\newcommand{\psiprime}{{\psi({\rm 2S})}}

\newcommand{\pp}{{\rm pp}}

\newcommand{\pt}{{p_{\rm T}}}
\newcommand{\aeff}{{A\epsilon}}

\newcommand{\mumu}{{\mu^+\mu^-}}
\newcommand{\invmass}{{M_{\mu\mu}}}
\newcommand{\lint}{{L_{\rm int}}}
\newcommand{\br}{{\rm BR}}

\newcommand{\ups}{{\rm\Upsilon}}
\newcommand{\stat}{{\rm (stat)}}
\newcommand{\syst}{{\rm (syst)}}

\graphicspath{{Figures/}}

\begin{document}

\begin{titlepage}
\title{Inclusive quarkonium production at forward rapidity\\in pp collisions at $\mathbf{\sqrt{s}=8}~$TeV}

\ShortTitle{Quarkonium measurement at $\sqrt{s}=8$ TeV} 

\Collaboration{ALICE Collaboration}
\ShortAuthor{ALICE Collaboration}   
\date{Received: date / Accepted: date}

\begin{abstract}
We report on the inclusive production cross sections of $\jpsi$, $\psiprime$, $\ups$(1S), $\ups$(2S) and $\ups$(3S), measured at forward rapidity with the ALICE detector in $\pp$ collisions at a center-of-mass energy $\sqrt{s}=8$~TeV. 
The analysis is based on data collected at the LHC and corresponds to an integrated luminosity of $1.23$ pb$^{-1}$. 
Quarkonia are reconstructed in the dimuon-decay channel.
The differential production cross sections are measured as a function of the transverse momentum $\pt$ and rapidity $y$, over the $\pt$ 
ranges $0<\pt<20$~GeV/$c$ for $\jpsi$, $0<\pt<12$~GeV/$c$ for all other resonances, and for $2.5<y<4$. The cross sections, integrated over $\pt$ and $y$, and assuming unpolarized quarkonia, are 
$\sigma_{\jpsi} = 8.98\pm0.04\pm0.82$~$\mu$b, 
$\sigma_{\psiprime} = 1.23\pm0.08\pm0.22$~$\mu$b, 
$\sigma_{\ups{\rm(1S)}} = 71\pm6\pm7$~nb, 
$\sigma_{\ups{\rm(2S)}} = 26\pm5\pm4$~nb and 
$\sigma_{\ups{\rm(3S)}} = 9\pm4\pm1$~nb, 
where the first uncertainty is statistical and the second one is systematic. 
These values agree, within at most $1.4\sigma$, with measurements performed by the LHCb collaboration in the same rapidity range.

\end{abstract}
\end{titlepage}
\setcounter{page}{2}

\section{\label{sec_introduction}Introduction}
The hadronic production of quarkonia, bound states of either a charm and anti-charm quark pair (e.g. $\jpsi$ and $\psiprime$) or a bottom and anti-bottom quark pair (e.g. $\ups$(1S), $\ups$(2S) and $\ups$(3S)), is generally understood as the result of a hard scattering that produces the heavy-quark pair, followed by the evolution of this pair into a colorless bound state. There are mainly three approaches used to describe quarkonium production, which differ mostly in the way the produced heavy-quark pair evolves into the bound state: the Color Evaporation Model~\cite{Fritzsch:1977ay,Amundson:1996qr}, the Color Singlet Model~\cite{Baier:1981uk} and Non-Relativistic QCD~\cite{Bodwin:1994jh}. To date, none of these approaches is able to describe consistently all data available on quarkonium production~\cite{Brambilla:2010cs,Andronic:2015wma}.

In this paper we present the production cross sections of $\jpsi$, $\psiprime$, $\ups$(1S), $\ups$(2S) and $\ups$(3S) at forward rapidity ($2.5<y<4$), measured in pp collisions at a center-of-mass energy $\sqrt{s}=8$~TeV with the ALICE detector. All quarkonia are reconstructed in the dimuon-decay channel. The differential production cross sections are measured as a function of the transverse momentum $\pt$ and rapidity $y$, over the $\pt$ 
ranges $0<\pt<20$~GeV/$c$ for $\jpsi$, $0<\pt<12$~GeV/$c$ for all other resonances, and for $2.5<y<4$. Our measurement extends the transverse momentum reach of the $\jpsi$ cross section from $\pt=12$~GeV/$c$ up to $\pt = 20$~GeV/$c$ with respect to results from LHCb~\cite{Aaij:2013yaa}. 
The $\psiprime$ results are the first published at this energy. For $\ups$ mesons, differential cross sections at forward rapidity and $\sqrt{s}=8$~TeV have already been published by LHCb~\cite{Aaij:2015awa}. Our measurement provides a unique cross-check of these results. Moreover, it is the first time ALICE measures  the $\ups$(3S) cross section.
All cross sections reported here are inclusive and contain, on top of the direct production of the quarkonium, a contribution from the decay of higher-mass excited states. Charmonium ($\jpsi$ and $\psiprime$) cross sections also contain a contribution from $b$-hadron decay.

The paper is organized as follows: the ALICE detector and the data sample used for this analysis are briefly described in Section~\ref{sec_detector}, the analysis procedure is discussed in Section~\ref{sec_analysis} and results are presented in Section~\ref{sec_results}.

\section{\label{sec_detector}Detector and data sample}
The ALICE detector is described in~\cite{ALICE} and its performance in~\cite{Abelev:2014ffa}. The following subsystems are used for measuring the quarkonium production cross sections at forward rapidity: the Muon Spectrometer~\cite{Aamodt:2011gj}, the first two layers of the Inner Tracking System (ITS)~\cite{ALICE_ITS}, the V0 scintillator hodoscopes~\cite{Abbas:2013taa} and the T0 Cherenkov counters~\cite{Bondila:2005xy}. 

The Muon Spectrometer consists of five tracking stations (MCH) comprising two planes of Cathode Pad Chambers each, followed by two trigger stations (MTR) consisting of two planes of Resistive Plate Chambers each.
It is used to detect muons produced in the pseudo-rapidity range $-4<\eta<-2.5$~\footnote{In the ALICE reference frame the muon spectrometer covers negative $\eta$. However, we use positive values when referring to the quarkonium rapidity $y$.}. The third tracking station is located inside a warm  $3$~$\si{\tesla\meter}$ dipole magnet, to allow for momentum measurements. This apparatus is completed by two absorbers that filter out hadrons and low $\pt$ muons, positioned (i) between the Interaction Point (IP) and the first tracking station, and (ii) between the last tracking station and the first trigger station. A third absorber, surrounding the beam pipe, protects the detectors from secondary particles produced inside the beam pipe. The MTR system delivers single- or di-muon triggers, of either same or opposite sign, with a programmable threshold on the transverse momentum of each muon. The ITS consists of 6 layers of silicon detectors, placed at radii ranging from $3.9$ to $43$~cm from the beam axis. Its two innermost layers are equipped with Silicon Pixel Detectors (SPD)
and cover the pseudo-rapidity ranges $|\eta|<2$ and $|\eta|<1.4$ for the inner and the outer layer, respectively. 
They are used for the reconstruction of the collision primary vertex. 
The V0 detectors are two scintillator arrays located on both sides of the IP and covering the pseudo-rapidity ranges $-3.7<\eta<-1.7$ and $2.8<\eta<5.1$. The T0 detectors are two arrays of quartz Cherenkov counters, also placed at forward rapidity on both sides of the IP and covering the pseudo-rapidity ranges $-3.3<\eta<-3$ and $4.6<\eta<4.9$. The coincidence of a signal in both sides of either the T0 or the V0 detectors is used as an interaction trigger and as
input for the luminosity determination.

The data used for this analysis have been collected in 2012. About 1400 proton bunches were circulating in each LHC beam. Collisions were delivered in a so-called beam-satellite mode, for which the high-intensity bunches of one of the two beams were collided with nearly-empty satellite bunches from the other~\cite{Abelev:2014ffa}. In this configuration, the average instantaneous luminosity delivered by the LHC to ALICE was about $5\times10^{30}$~cm$^{-2}$s$^{-1}$. 
The number of interactions per bunch-satellite crossing was about 0.01 on average with a corresponding pile-up probability of about $0.5$\%, reaching a maximum of $\sim1$\%.

Events are selected using a dimuon trigger which requires that two muons of opposite sign are detected in the MTR, with a threshold of $1$~GeV/$c$ applied online to the $\pt$ of each muon, in coincidence with the crossing of two bunches at the IP. 
The data sample recorded with this trigger corresponds to an integrated luminosity $\lint = 1.23$~pb$^{-1}$. It is evaluated on a run-by-run basis by multiplying the dimuon trigger live-time with the delivered luminosity. The latter is estimated using the number of T0-based trigger counts and the corresponding cross section, $\sigma_{\rm T0}$, measured using the van der Meer scan method~\cite{vanderMeer:1968zz}. 
The systematic uncertainty on this quantity includes contributions from (i) the measurement of $\sigma_{\rm T0}$ itself and (ii) the difference between the luminosity measured with the T0 detectors and the one measured with the V0 detectors.
The quadratic sum of these contributions amounts to about $5$\% and is correlated between all measurements presented in this paper.

\section{\label{sec_analysis}Analysis}
The differential quarkonium production cross section in a given $\pt$ and $y$ interval is:
\begin{equation}
\label{eq_cross_section}
\frac{d^2\sigma}{d\pt dy}=\frac{1}{\Delta\pt\Delta y}\frac{1}{\lint}\frac{N}{\br_{\mu\mu}\aeff},
\end{equation}
where $\br_{\mu\mu}$ is the branching ratio of the quarkonium state in two muons, 
$\Delta\pt$ and $\Delta y$ are the widths of the $\pt$ and $y$ intervals under consideration, $N$ is the measured number of quarkonia in these intervals and $\aeff$ is the product of the corresponding acceptance and efficiency corrections, which account for detector effects and analysis cuts. 
The branching ratio values and uncertainties have been taken from the Particle Data Group (PDG)~\cite{Agashe:2014kda}.
The other ingredients, namely $N$ and $\aeff$, have been evaluated using the analysis procedure described in~\cite{Abelev:2014qha}.

The number of quarkonia measured in a given $\pt$ and $y$ interval is evaluated using fits to the invariant mass distribution of opposite-sign muon pairs $\mumu$. These pairs are formed by combining the tracks reconstructed in the muon spectrometer and selected using the same criteria as in~\cite{Abelev:2014qha}:
\begin{itemize}
\item muon identification is performed by matching each track reconstructed in the MCH with a track in the MTR that fulfills the trigger condition;
\item tracks are selected in the pseudo-rapidity range $-4<\eta<-2.5$, which corresponds to the muon spectrometer geometrical acceptance;
\item the transverse position of the tracks at the end of the front absorber, $R_{\rm{abs}}$, is in the range $17.6 < R_{\rm{abs}} < 89.5$~cm, in order to reject muons crossing the high-density section of the front absorber;
\item tracks must pass a cut on the product of their total momentum, $p$, and their distance to the primary vertex in the transverse plane, called DCA. The maximum value allowed is set to $6\times\sigma_{\rm pDCA}$, where $\sigma_{\rm pDCA}$ is the resolution on this quantity, which accounts for the total momentum and angular resolutions of the muon spectrometer as well as for the multiple scattering in the front absorber. This cut reduces the contamination of fake tracks and particles from beam-gas interactions.
\end{itemize}

The fit to the $\mumu$ invariant mass distribution is performed separately in the charmonium and bottomonium regions, and for each $\pt$ and $y$ interval under consideration. In all cases the fitting function consists of a background to which two (three) signal functions are added, one per charmonium (bottomonium) state under study.

\begin{figure}[h]
\begin{center}
\begin{tabular}{cc}
\includegraphics[width=0.48\linewidth,keepaspectratio]{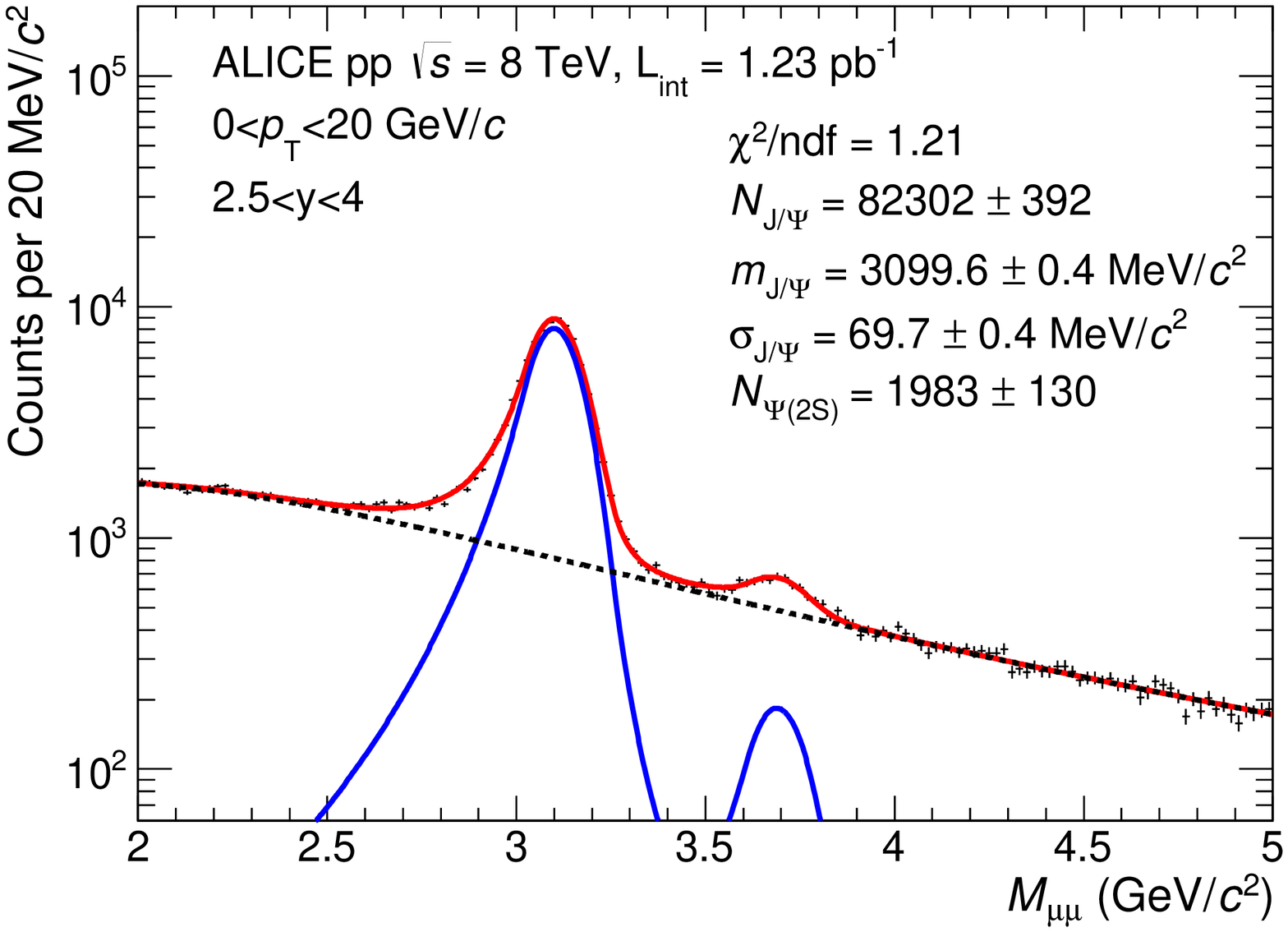}&
\includegraphics[width=0.48\linewidth,keepaspectratio]{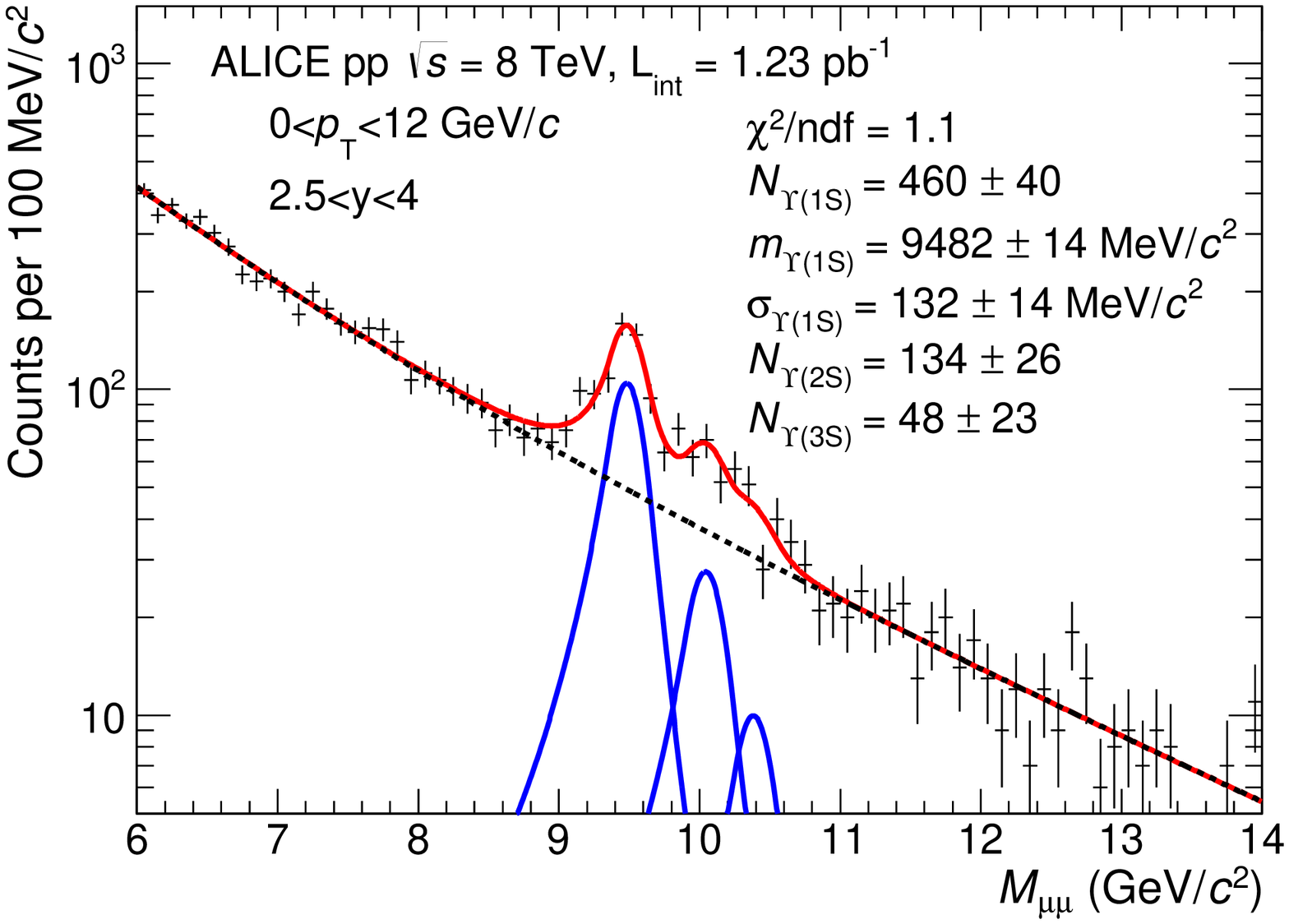}\\
\end{tabular}
\end{center}
\caption{\label{figure_signal} Dimuon invariant mass distributions in the region of charmonia (left) and bottomonia (right). Dashed lines correspond to the background. Solid lines correspond to either the signal functions, or the sum of all signal and background functions. In the charmonia region, the sum of two extended Crystal Ball functions is used for the signal and a pseudo-Gaussian function is used for the background. In the bottomonia region, the sum of three extended Crystal Ball functions is used for the signal and the sum of two exponential functions is used for the background.}
\end{figure}

For charmonia, the fit is performed over the invariant mass range $2<\invmass<5$~GeV/$c^2$. For the background component, either a pseudo-Gaussian function whose width varies linearly with the invariant mass or the product of an exponential function and a fourth order polynomial function have been used, with all parameters left free in the fit. For the signal, the sum of either two extended Crystal Ball functions (one for each resonance) or two pseudo-Gaussian functions have been used~\cite{ALICE-PUBLIC-2015-006}. Both functions (Crystal Ball or pseudo-Gaussian) consist of a Gaussian core, to which parametrized tails are added on both sides, which fall off slower than for a Gaussian function. Due to the poor signal-to-background (S/B) ratio in the tail regions, the values of the parameters that enter the definition of these tails have been evaluated using Monte Carlo (MC) simulations described later in this Section, and kept fixed in the fit. The $\jpsi$ and $\psiprime$ signals are fitted simultaneously. For the $\jpsi$, the mass, width and normalization of the signal function are left free. For the $\psiprime$, only the normalization is free, whereas the mass and the width are calculated from the values obtained for the $\jpsi$: the mass is computed so that the difference with respect to the $\jpsi$ mass is the same as quoted by the PDG~\cite{Agashe:2014kda}; the width is derived from the $\jpsi$ width using a scale factor of about 1.1, estimated in MC simulations and validated with fits to the $\pt$- and $y$-integrated invariant mass distributions from the data, with both widths left free. An example of fit to the $\pt$- and $y$-integrated dimuon invariant mass distribution in the $\jpsi$ and $\psiprime$ mass region is shown in the left panel of Fig.~\ref{figure_signal}. The result from this fit is used for the computation of the charmonium cross sections quoted at the beginning of section~\ref{sec_results}.

For the $\ups$ resonances, the fit is performed over the invariant mass range $6<\invmass<14$~GeV/$c^2$. The same signal functions as for the $\jpsi$ and $\psiprime$ have been used for each of the three resonances, albeit with different values for the parameters of the tails. For the background component, either the sum of two exponential functions or the sum of two power law functions have been used, with all parameters left free. The masses and widths of the $\ups$(2S) and $\ups$(3S) resonances have been fixed to the ones of the $\ups$(1S) in a similar way as for the $\psiprime$ and $\jpsi$ case, and using a similar scale factor for the width. An example of fit to the $\pt$- and $y$-integrated dimuon invariant mass distribution in the $\ups$ mass region is shown in the right panel of Fig.~\ref{figure_signal}.

The number of quarkonia is taken as the mean of the values obtained when (i) combining all possible signal and background functions described above; (ii) varying the parameters that have been fixed, such as those of the tails of the signal functions or the ratio between the $\psiprime$ and the $\jpsi$ signal widths, and (iii) modifying the mass range used for the fit.

Approximately 82500 $\jpsi$, 1850 $\psiprime$, 480 $\ups$(1S), 140~$\ups$(2S) and 50~$\ups$(3S) are measured. The corresponding S/B ratios, evaluated 
within three times the width of the signal function with respect to the quarkonium mass
are $4.5$ for $\jpsi$, $0.2$ for $\psiprime$, $1$ for $\ups$(1S), $0.4$ for $\ups$(2S) and $0.2$ for $\ups$(3S).
This statistics allows us to divide the data sample further as a function of either $\pt$ or $y$ for $\jpsi$, $\psiprime$ and $\ups$(1S). For $\ups$(2S), only two bins in $y$ are measured, whereas for $\ups$(3S), only the $\pt$- and $y$-integrated value is provided, due to limited statistics. For $\jpsi$, the S/B ratio increases from 3 to 10 with increasing $\pt$ and from $4$ to $6$ with increasing $y$. For $\psiprime$, it increases from $0.1$ to $0.9$ with increasing $\pt$ and from $0.1$ to $0.2$ with increasing $y$. For $\ups$(1S), it increases from $0.8$ to $1.4$ with increasing $\pt$ and shows no significant variation with respect to $y$. No significant variation with respect to $y$ is observed for $\ups$(2S) either. 

The systematic uncertainty on the signal extraction is estimated by taking the root mean square of the values from which the number of quarkonia is derived.
For a given quarkonium state, this uncertainty is considered as uncorrelated as a function of both $\pt$ and $y$. It is however partially correlated between $\jpsi$ and $\psiprime$ as well as among the three resonances of the $\ups$ family. For $\jpsi$ this uncertainty increases from less than $1$\% to $14$\% with increasing $\pt$. It shows no significant variation with respect to $y$ and amounts to about $1$\%. Larger values are obtained for $\psiprime$ due to the smaller S/B ratio. For instance, the uncertainty reaches $18$\% in the $y$ interval $2.5<y<2.75$. In the $\ups$ sector, the systematic uncertainty is about $3$\%, $6$\% and $10$\% for $\ups$(1S), $\ups$(2S) and $\ups$(3S), respectively, with little variation as a function of either $\pt$ or $y$.

Acceptance and efficiency corrections, $\aeff$, are evaluated separately for each quarkonium state using MC simulations. Each state is generated randomly using realistic $\pt$ and $y$ probability distribution functions~\cite{Aamodt:2011gj,Abelev:2014qha}. It is decayed in two muons, properly accounting for the possible emission of an accompanying radiative photon~\cite{Lange:2001uf,Barberio:1990ms}. The muons are then tracked in a model of the apparatus obtained with GEANT 3.21~\cite{GEANT3} which includes a realistic description of the detector performance during data taking as well as its variation with time. The same procedure and analysis cuts as for data are then applied to the MC simulations for track reconstruction and measurement of the quarkonium yields. All simulated quarkonia are assumed to be unpolarized, consistently with existing measurements~\cite{alice:2012pol,Aaij:2013nlm,Chatrchyan:2013cla,Chatrchyan:2012woa}.

The systematic uncertainty on $\aeff$ has several contributions: (i) the parametrization of the input $\pt$ and $y$ distributions; (ii) the track reconstruction efficiency and the accuracy with which the detector performance is reproduced in the MC simulations; (iii) the trigger efficiency and (iv) the matching between tracks reconstructed in the MCH and tracks reconstructed in the MTR. These contributions have been evaluated using the same procedures as in~\cite{Abelev:2014qha}, for the first one by utilizing several alternative input $\pt$ and $y$ distributions, and for the other three by comparing data and MC at the single muon level and propagating the resulting differences to the dimuon case. The resulting systematic uncertainty is the quadratic sum of these contributions. It is partially correlated as a function of both $\pt$ and $y$. For all quarkonium states, it amounts to about $8$\% on average, increases from $7$\% to $9$\% with increasing $\pt$ and shows no visible dependence on $y$.

An additional correction is applied to the number of measured quarkonia, to account for the observation that a fraction of the opposite-sign muon pairs of a given quarkonium state is sometimes misidentified by the trigger system as a same-sign pair and thus missed. The magnitude of this effect could not be properly reproduced in the MC simulations and is therefore not accounted for in the $\aeff$ corrections. For $\jpsi$ and $\ups$(1S), it is instead evaluated directly on data by means of a dedicated trigger configuration that selects both same- and opposite-sign muon pairs instead of opposite-sign pairs only. The statistical and systematic uncertainties on the extraction of the signal in each configuration are used to evaluate the systematic uncertainty on the resulting correction. For $\jpsi$, the correction amounts to about $1$\% on the $\pt$- and $y$-integrated yield. It increases from $0.6$\% to $8$\% with increasing $\pt$ and shows little dependence on $y$. Slightly larger values are obtained for $\ups$(1S) albeit with larger uncertainties. For $\psiprime$, the same corrections as for $\jpsi$ have been used, whereas for $\ups$(2S) and $\ups$(3S) we used the same corrections as for $\ups$(1S). 

Tables~\ref{tab_syst_charm} and~\ref{tab_syst_bottom} provide a summary of the relative systematic uncertainties on the charmonia and bottomonia cross sections, respectively.

\begin{table}
\begin{center}
\begin{tabular}{c|c|c}
Source & $\jpsi$ & $\psiprime$ \\
\hline
Luminosity&$5$\%&$5$\% \\
Branching ratio & $<1$\% & $11$\% \\
Signal extraction & $1$\% ($<1$\%-$14$\%) & $10$\% ($6$\%-$18$\%) \\
Acceptance$\times$efficiency & $8$\% ($7$\%-$9$\%) & $8$\% ($7$\%-$9$\%) \\
Trigger sign & $<1$\% ($<1$\%-$3$\%) & $<1$\% ($<1$\%-$3$\%) \\
\end{tabular}
\end{center}
\caption{\label{tab_syst_charm}Relative systematic uncertainties associated to the $\jpsi$ and $\psiprime$ cross section measurements. Values in parenthesis correspond to minimum and maximum values as a function of $\pt$ and $y$.}
\end{table}

\begin{table}
\begin{center}
\begin{tabular}{c|c|c|c}
Source & $\ups$(1S) & $\ups$(2S) & $\ups$(3S) \\
\hline
Luminosity & $5$\% & $5$\% & $5$\% \\
Branching ratio & $2$\% & $9$\% & $10$\% \\
Signal extraction & $3$\% ($2$\%-$6$\%) & $6$\% ($5$\%-$9$\%) & $10$\% \\
Acceptance$\times$efficiency& $8$\% ($7$\%-$9$\%) & $8$\% & $8$\% \\
Trigger sign&$1$\% ($1$\%-$5$\%)&$1$\% ($1$\%-$2$\%)&$1$\%\\
\end{tabular}
\end{center}
\caption{\label{tab_syst_bottom}Relative systematic uncertainties associated to the $\ups$(1S), $\ups$(2S) and $\ups$(3S) cross section measurements. Values in parenthesis correspond to minimum and maximum values as a function of $\pt$ and $y$.}
\end{table}

\section{\label{sec_results}Results}

The measured inclusive quarkonium production cross sections, integrated over $0<\pt<20$~GeV/$c$ for $\jpsi$, $0<\pt<12$~GeV/$c$ for all other resonances, and $2.5<y<4$, are:

$\sigma_{\jpsi} = 8.98\pm0.04\stat\pm0.82\syst$~$\mu$b, 

$\sigma_{\psiprime} = 1.23\pm0.08\stat\pm0.22\syst$~$\mu$b, 

$\sigma_{\ups{\rm(1S)}} = 71\pm6\stat\pm7\syst$~nb, 

$\sigma_{\ups{\rm(2S)}} = 26\pm5\stat\pm4\syst$~nb and 

$\sigma_{\ups{\rm(3S)}} = 9\pm4\stat\pm1\syst$~nb.

These values are in agreement, within at most $1.4\sigma$, with measurements performed by LHCb at the same energy and in the same rapidity range~\cite{Aaij:2013yaa,Aaij:2015awa}, assuming that all uncertainties but the one on the branching ratios are uncorrelated between the two experiments. 
For $\jpsi$, our cross section value corresponds to an increase of $(29\pm17)$\% with respect to the ALICE measurement at $\sqrt{s}=7$~TeV~\cite{Abelev:2014qha}. A similar increase is observed for $\psiprime$ and for the $\ups$ resonances, albeit with larger uncertainties.

\begin{figure}[h]
\begin{center}
\begin{tabular}{cc}
\includegraphics[width=0.48\linewidth,keepaspectratio]{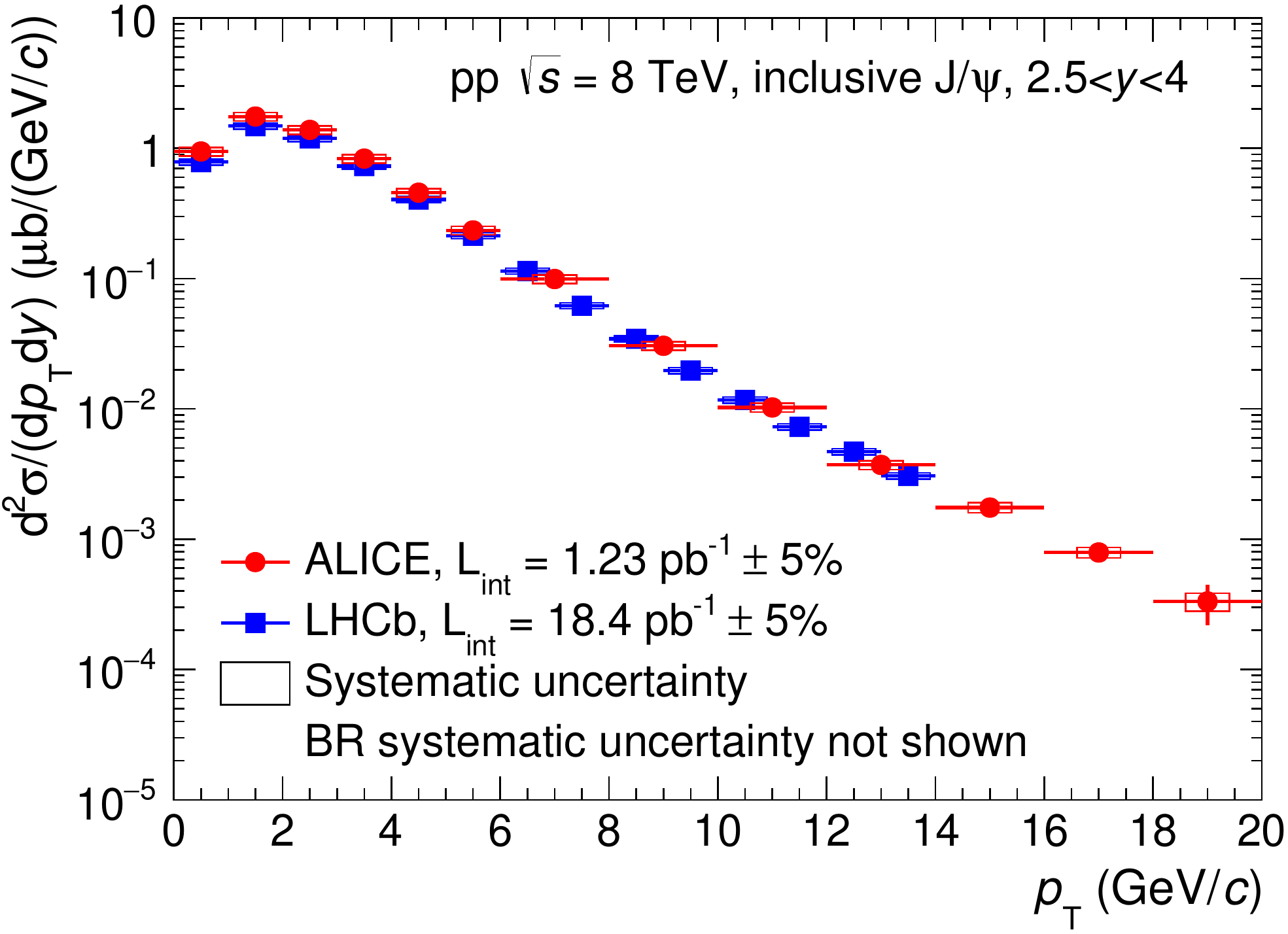}&
\includegraphics[width=0.48\linewidth,keepaspectratio]{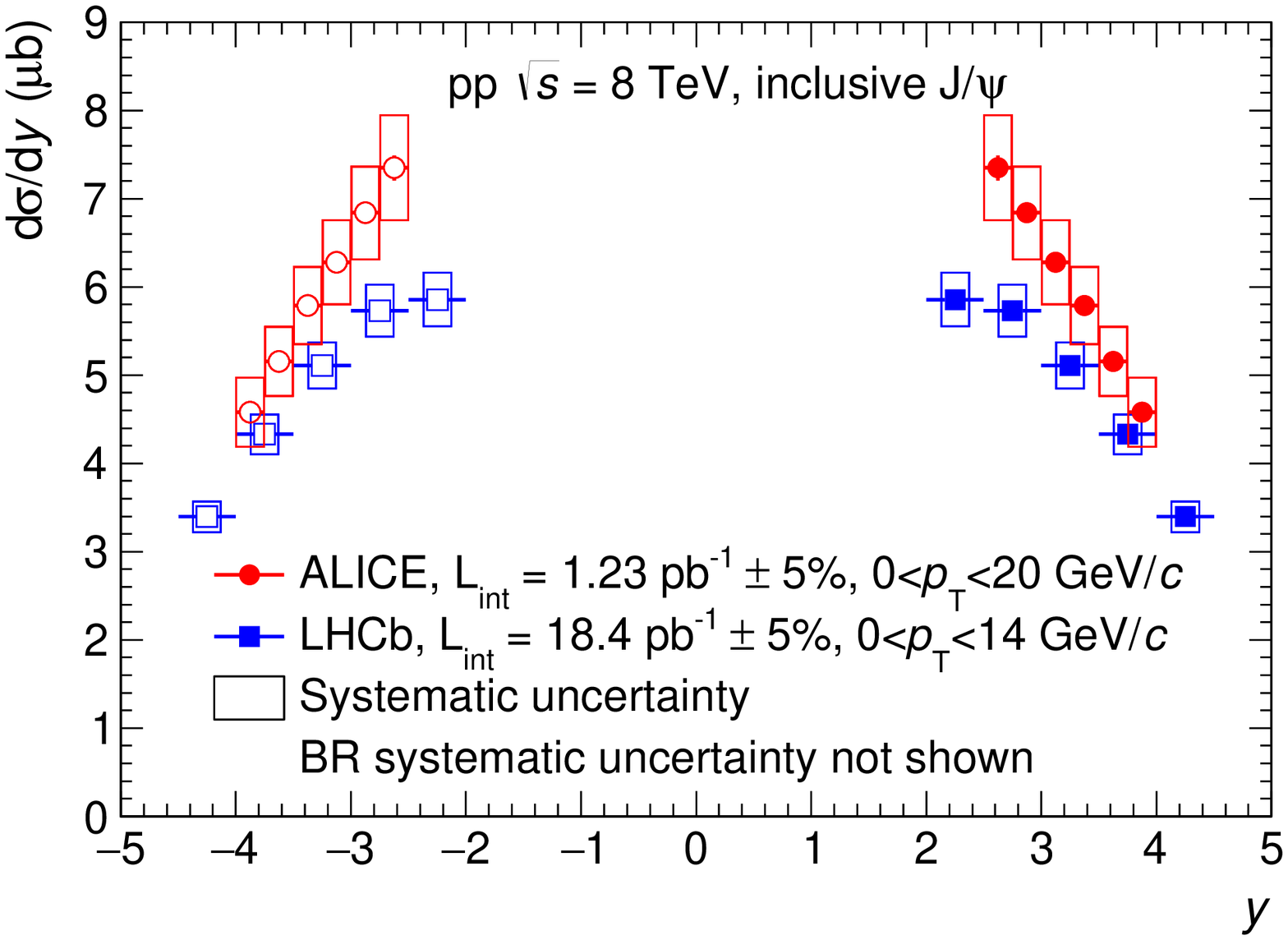}\\
\includegraphics[width=0.48\linewidth,keepaspectratio]{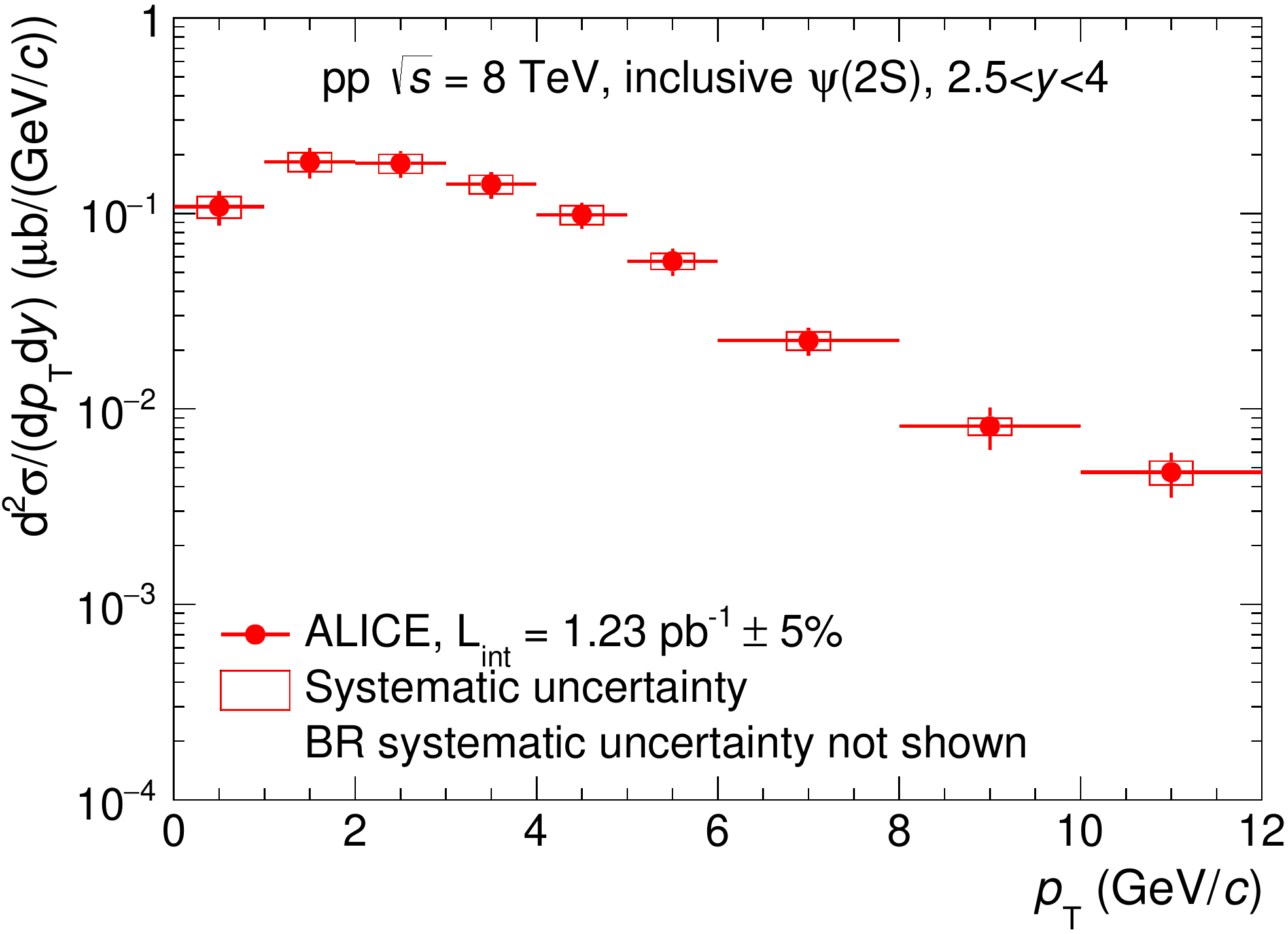}&
\includegraphics[width=0.48\linewidth,keepaspectratio]{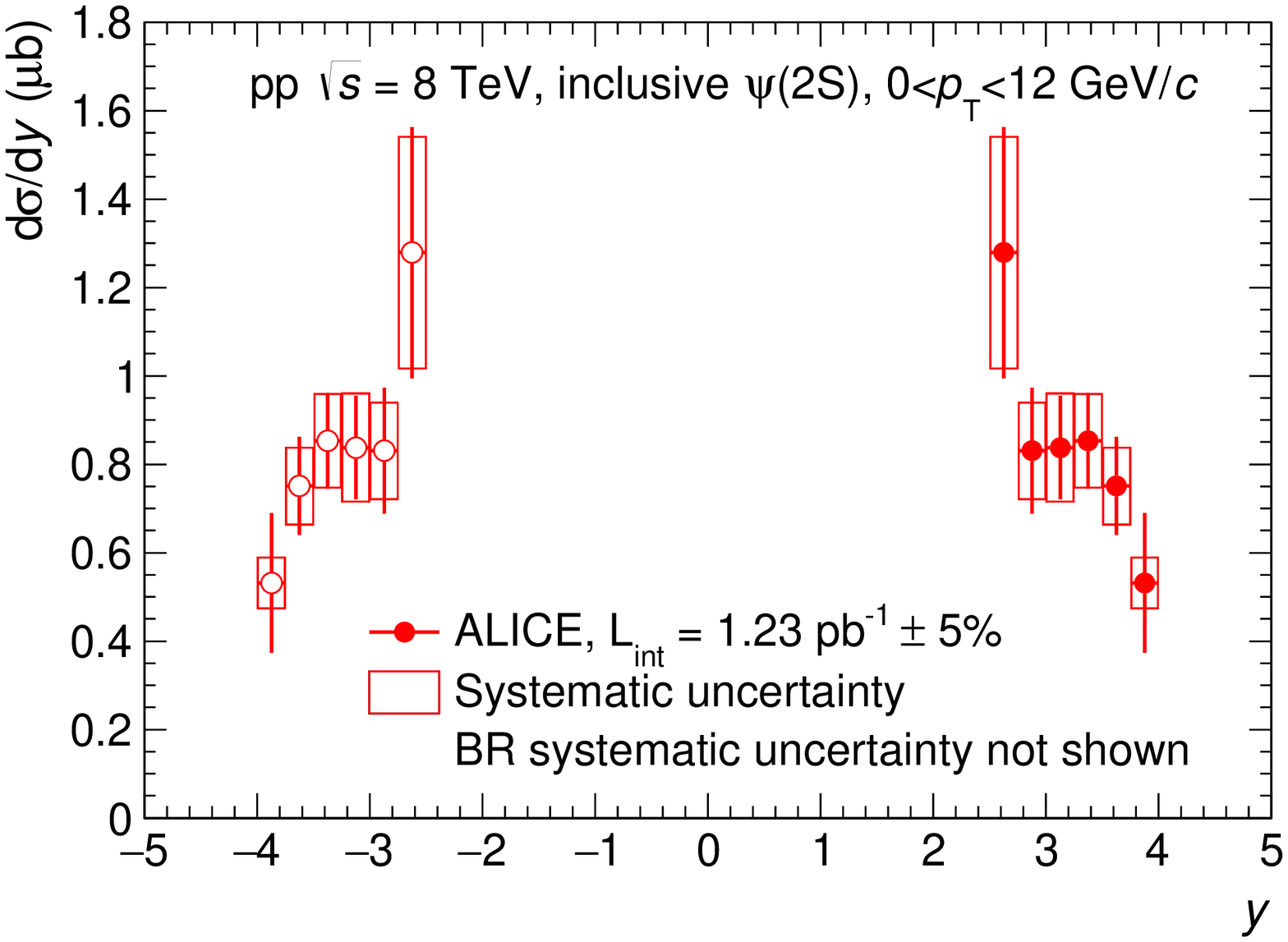}\\
\end{tabular}
\end{center}
\caption{\label{figure_charmonia} $\jpsi$ (top) and $\psiprime$ (bottom) differential cross sections as a function of $\pt$ (left) and $y$ (right). $\jpsi$ results are compared to LHCb measurement at $\sqrt{s}=8$~TeV~\cite{Aaij:2013yaa}. Open symbols are the reflection of the positive-$y$ measurements with respect to $y=0$. Vertical error bars are the statistical uncertainties. Boxes are the systematic uncertainties. Branching ratio uncertainties are not included.}
\end{figure} 

Figure~\ref{figure_charmonia} shows the inclusive differential production cross sections of $\jpsi$ (top) and $\psiprime$ (bottom) as a function of $\pt$ (left) and $y$ (right) in pp collisions at $\sqrt{s}=8$~TeV. In all the plots, the error bars represent the statistical uncertainties and the boxes correspond to the systematic uncertainties. Branching ratio uncertainties are not included. The $\jpsi$ $\pt$- and $y$-differential cross sections are compared to measurements by LHCb at the same energy~\cite{Aaij:2013yaa}. 
The quoted LHCb values correspond to the sum of the prompt and $b$-meson decay contributions to the $\jpsi$ production. 
For the comparison as a function of $\pt$, the provided double-differential ($\pt$ and $y$) values have been re-summed to match ALICE $y$ coverage.
A reasonable agreement is observed between the two experiments. Although the ALICE measurements are systematically above those of LHCb especially at low $\pt$ and small $|y|$, in both cases the differences do not exceed $1.7\sigma$.
The ALICE measurement extends the $\pt$ reach of the $\jpsi$ cross section from $14$~GeV/$c$ to $20$~GeV/$c$ with respect to published results. The $\psiprime$ cross sections constitute the first measurement performed at this energy. 

\begin{figure}[h]
\begin{center}
\begin{tabular}{cc}
\includegraphics[width=0.48\linewidth,keepaspectratio]{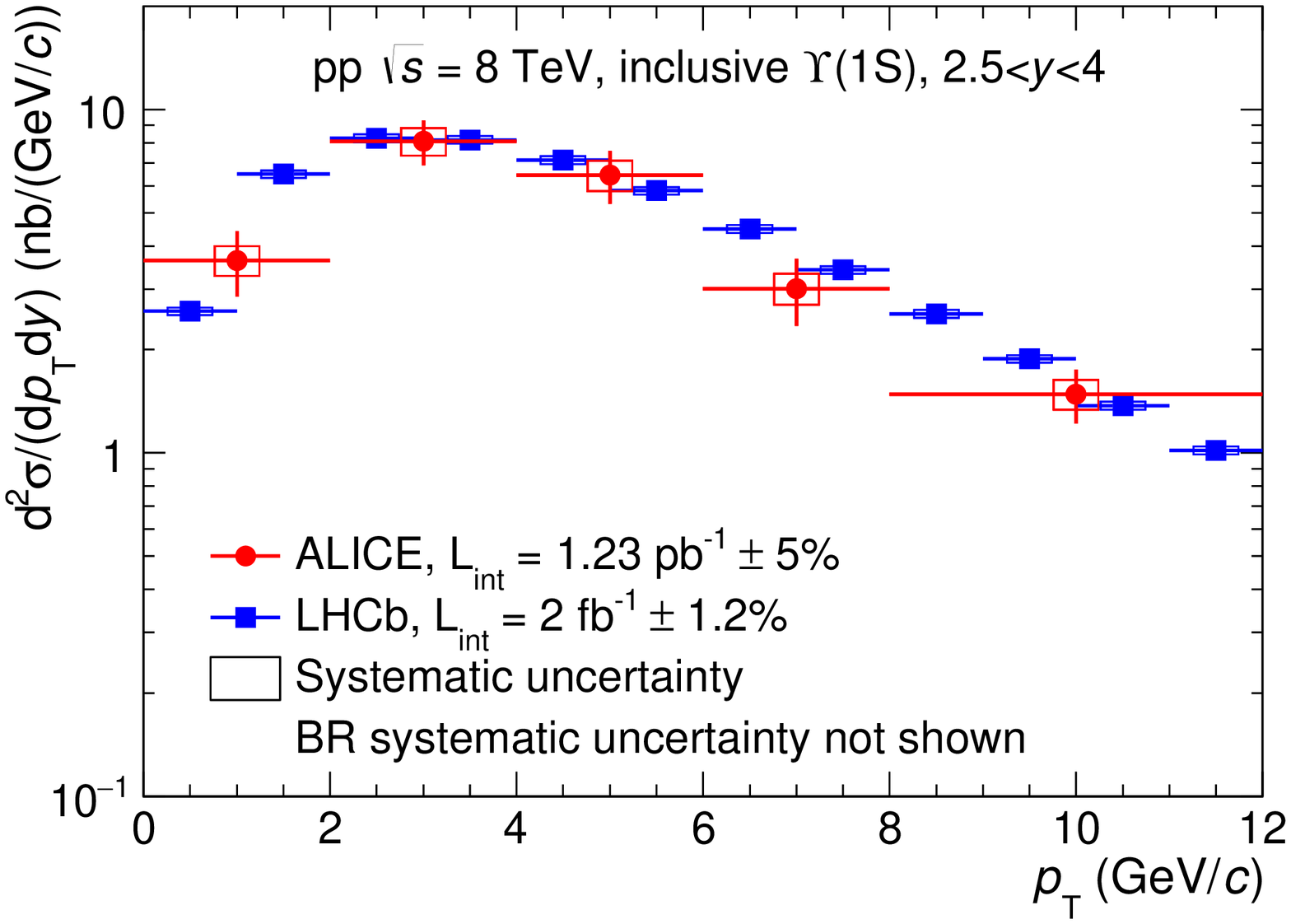}&
\includegraphics[width=0.48\linewidth,keepaspectratio]{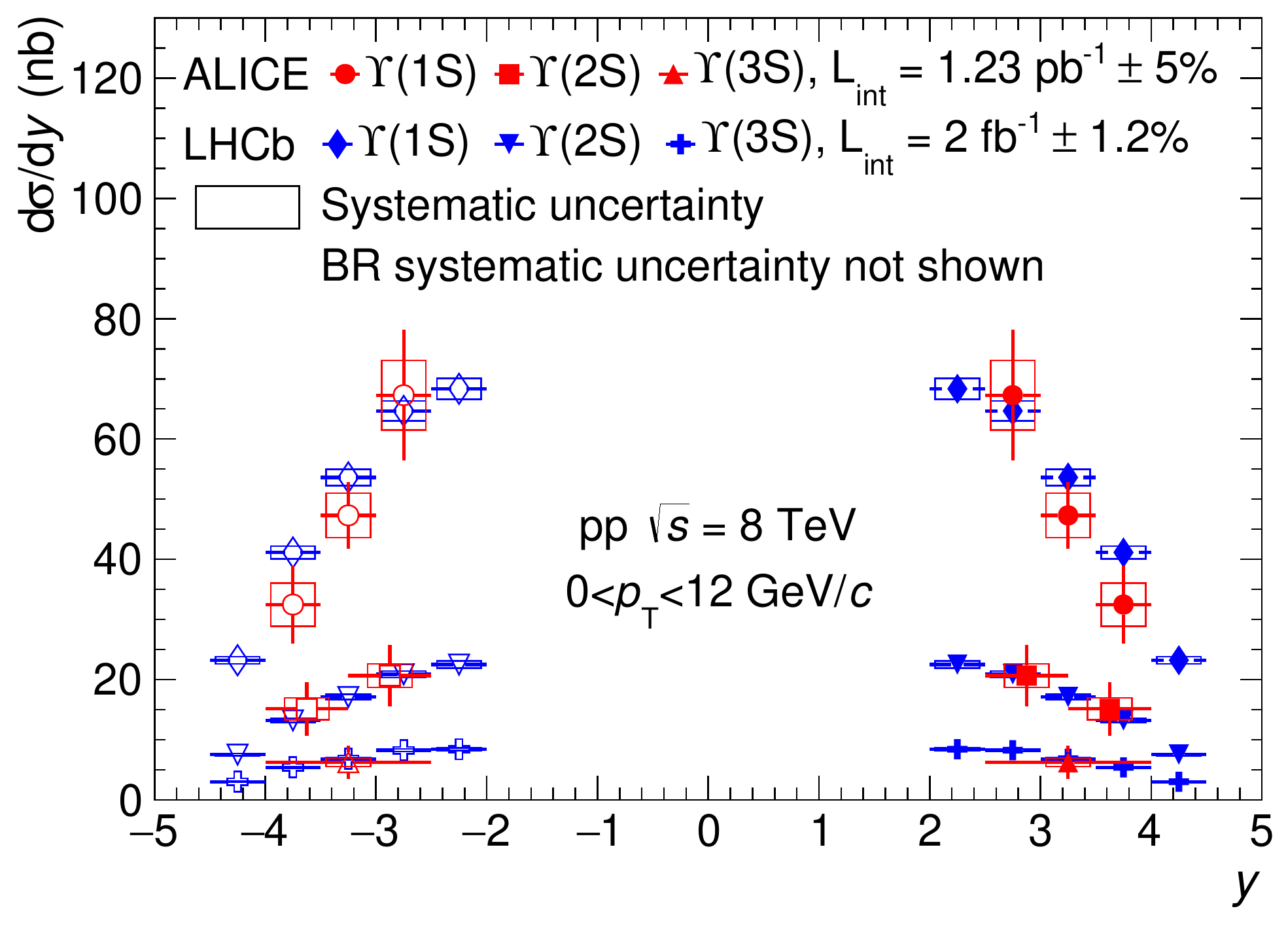}
\end{tabular}
\end{center}
\caption{\label{figure_bottomonia}Differential cross section of $\ups$(1S) as a function of $\pt$ (left) and differential cross sections of $\ups$(1S), $\ups$(2S) and $\ups$(3S) as a function of $y$ (right) measured by ALICE and LHCb~\cite{Aaij:2015awa}. Open symbols are the reflection of the positive-$y$ measurements with respect to $y=0$.}
\end{figure}

Figure~\ref{figure_bottomonia} shows the inclusive differential production cross sections of $\ups$(1S) as a function of $\pt$ (left) and of the $\ups$(1S), $\ups$(2S) and $\ups$(3S) as a function of $y$ (right). Results are compared to measurements by LHCb at the same energy~\cite{Aaij:2015awa}. For the comparison as a function of $\pt$ (resp. $y$), the double-differential values provided by LHCb have been re-summed to match the $y$ (resp. $\pt$) range of ALICE. Moreover, although the $\pt$ range measured by LHCb extends to values as large as $30$~GeV/$c$, we only show these measurements in the range $0<\pt<12$~GeV/$c$, which is more relevant for the comparison to our result. A reasonable agreement is observed between the two experiments. For $\ups$(1S), ALICE measurements are systematically lower than those from LHCb, however the differences do not exceed $1.2\sigma$ as a function of either $\pt$ or $y$.

The inclusive $\psiprime$-to-$\jpsi$ cross section ratio at $\sqrt{s}=8$~TeV, integrated over $\pt$ and $y$ is 
$\sigma_\psiprime / \sigma_\jpsi = 0.14\pm0.01\pm0.02$,
 the $\ups$(2S)-to-$\ups$(1S) ratio is $\sigma_{\ups\rm(2S)} / \sigma_{\ups\rm(1S)} = 0.37\pm0.08\pm0.04$ 
and the $\ups$(3S)-to-$\ups$(1S) ratio, $\sigma_{\ups\rm(3S)} / \sigma_{\ups\rm(1S)} = 0.12\pm0.05\pm0.02$, 
where the first uncertainty is statistical and the second one is systematic. When forming these ratios, the systematic uncertainty on the signal extraction is slightly reduced, due to correlations between the numerator and the denominator. All other sources of systematic uncertainties cancel, except for the uncertainties on the input $\pt$ and $y$ parametrizations in the MC, and on ${\rm BR}_{\mu\mu}$. The $\psiprime$-to-$\jpsi$ and $\ups$(2S)-to-$\ups$(1S) ratios are consistent with the values obtained in the same rapidity range at $\sqrt{s}=7$~TeV~\cite{Abelev:2014qha}.

\begin{figure}[h]
\begin{center}
\begin{tabular}{cc}
\includegraphics[width=0.48\linewidth,keepaspectratio]{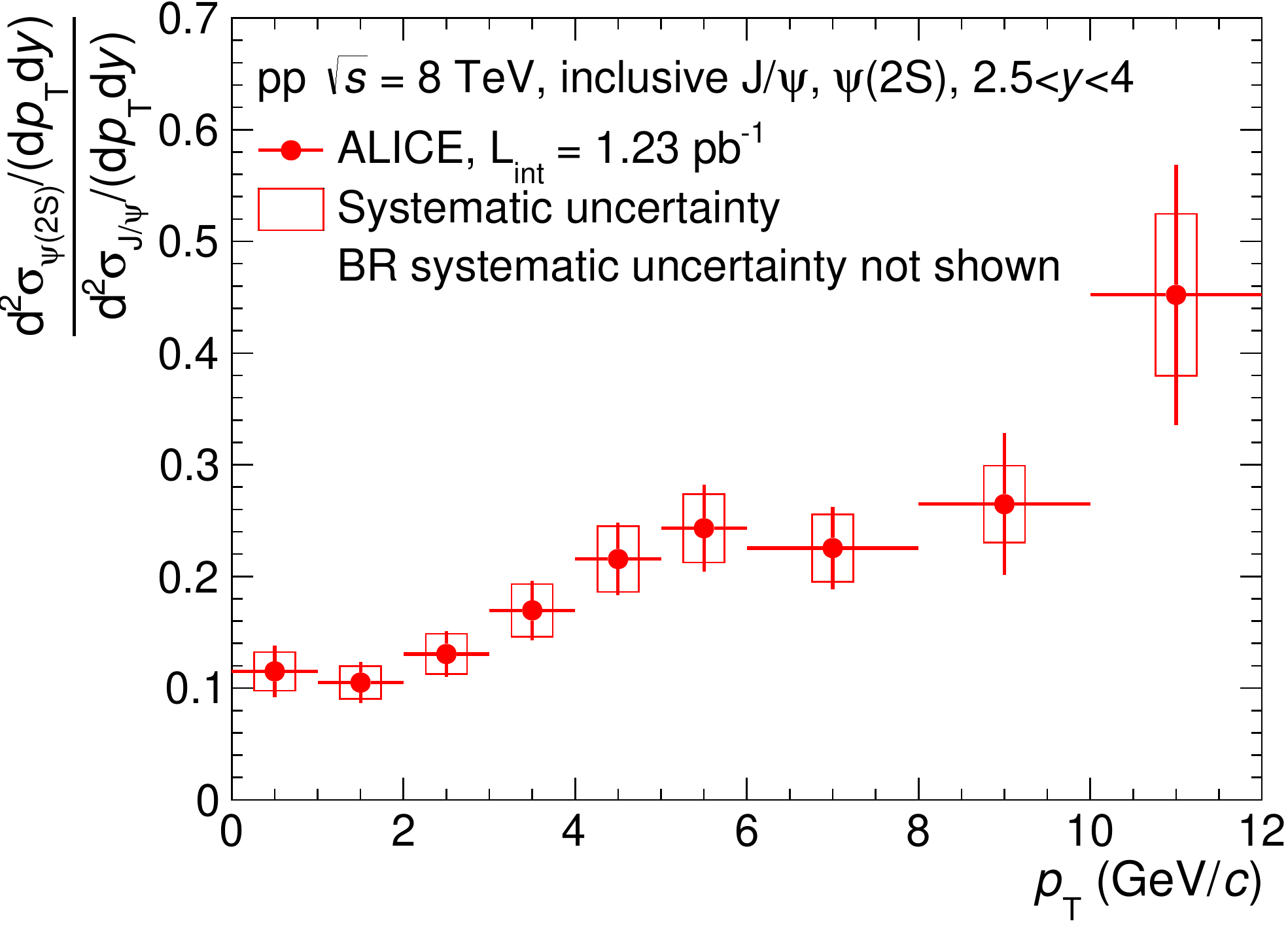}&
\includegraphics[width=0.48\linewidth,keepaspectratio]{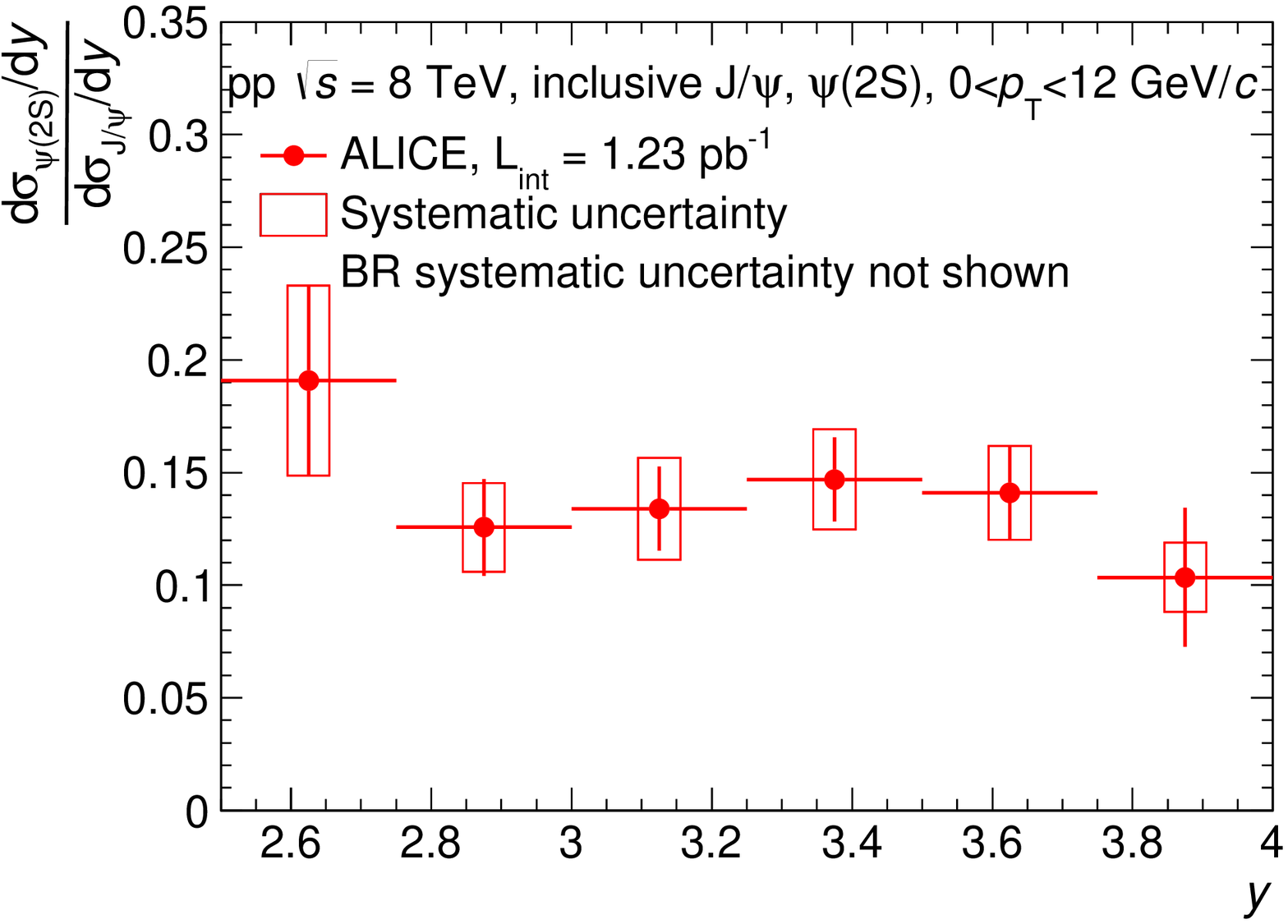}\\
\end{tabular}
\end{center}
\caption{\label{figure_charmonia_ratio}$\psiprime$-to-$\jpsi$ cross section ratio as a function of $\pt$ (left) and $y$ (right).}
\end{figure}

Figure~\ref{figure_charmonia_ratio} shows the $\psiprime$-to-$\jpsi$ cross section ratio as a function of $\pt$ (left) and $y$ (right). This ratio increases as a function of $\pt$ with a slope that is similar to the one measured at $\sqrt{s}=7$~TeV~\cite{Abelev:2014qha}. It shows no visible variation as a function of $y$, as was also the case at 7~TeV.

\section{\label{sec_conclusion}Conclusion}

The inclusive production cross section of $\jpsi$, $\psiprime$, $\ups$(1S), $\ups$(2S) and $\ups$(3S) as a function of $\pt$ and $y$ have been measured using the ALICE detector at forward rapidity ($2.5<y<4$) in pp collisions at $\sqrt{s}=8$~TeV. 
The $\jpsi$ cross section is larger by $(29\pm17)$\% than the one measured at $\sqrt{s}=7$~TeV~\cite{Abelev:2014qha}. A similar increase is observed for the other quarkonium states albeit with larger uncertainties.
The integrated results are in agreement within at most $1.4\sigma$ with measurements performed by LHCb in the same rapidity range.
For the differential measurements, differences with LHCb do not exceed $1.7\sigma$ for charmonia and $1.2\sigma$ for bottomonia.
These measurements provide a valuable cross-check of the already published results of the same quantities as well as additional experimental constraints on quarkonium production models.

\newenvironment{acknowledgement}{\relax}{\relax}
\begin{acknowledgement}
\section*{Acknowledgements}

The ALICE Collaboration would like to thank all its engineers and technicians for their invaluable contributions to the construction of the experiment and the CERN accelerator teams for the outstanding performance of the LHC complex.
The ALICE Collaboration gratefully acknowledges the resources and support provided by all Grid centres and the Worldwide LHC Computing Grid (WLCG) collaboration.
The ALICE Collaboration acknowledges the following funding agencies for their support in building and
running the ALICE detector:
State Committee of Science,  World Federation of Scientists (WFS)
and Swiss Fonds Kidagan, Armenia;
Conselho Nacional de Desenvolvimento Cient\'{\i}fico e Tecnol\'{o}gico (CNPq), Financiadora de Estudos e Projetos (FINEP),
Funda\c{c}\~{a}o de Amparo \`{a} Pesquisa do Estado de S\~{a}o Paulo (FAPESP);
National Natural Science Foundation of China (NSFC), the Chinese Ministry of Education (CMOE)
and the Ministry of Science and Technology of China (MSTC);
Ministry of Education and Youth of the Czech Republic;
Danish Natural Science Research Council, the Carlsberg Foundation and the Danish National Research Foundation;
The European Research Council under the European Community's Seventh Framework Programme;
Helsinki Institute of Physics and the Academy of Finland;
French CNRS-IN2P3, the `Region Pays de Loire', `Region Alsace', `Region Auvergne' and CEA, France;
German Bundesministerium fur Bildung, Wissenschaft, Forschung und Technologie (BMBF) and the Helmholtz Association;
General Secretariat for Research and Technology, Ministry of Development, Greece;
Hungarian Orszagos Tudomanyos Kutatasi Alappgrammok (OTKA) and National Office for Research and Technology (NKTH);
Department of Atomic Energy and Department of Science and Technology of the Government of India;
Istituto Nazionale di Fisica Nucleare (INFN) and Centro Fermi -
Museo Storico della Fisica e Centro Studi e Ricerche ``Enrico Fermi'', Italy;
MEXT Grant-in-Aid for Specially Promoted Research, Ja\-pan;
Joint Institute for Nuclear Research, Dubna;
National Research Foundation of Korea (NRF);
Consejo Nacional de Cienca y Tecnologia (CONACYT), Direccion General de Asuntos del Personal Academico(DGAPA), M\'{e}xico, Amerique Latine Formation academique - European Commission~(ALFA-EC) and the EPLANET Program~(European Particle Physics Latin American Network);
Stichting voor Fundamenteel Onderzoek der Materie (FOM) and the Nederlandse Organisatie voor Wetenschappelijk Onderzoek (NWO), Netherlands;
Research Council of Norway (NFR);
National Science Centre, Poland;
Ministry of National Education/Institute for Atomic Physics and National Council of Scientific Research in Higher Education~(CNCSI-UEFISCDI), Romania;
Ministry of Education and Science of Russian Federation, Russian
Academy of Sciences, Russian Federal Agency of Atomic Energy,
Russian Federal Agency for Science and Innovations and The Russian
Foundation for Basic Research;
Ministry of Education of Slovakia;
Department of Science and Technology, South Africa;
Centro de Investigaciones Energeticas, Medioambientales y Tecnologicas (CIEMAT), E-Infrastructure shared between Europe and Latin America (EELA), Ministerio de Econom\'{i}a y Competitividad (MINECO) of Spain, Xunta de Galicia (Conseller\'{\i}a de Educaci\'{o}n),
Centro de Aplicaciones Tecnológicas y Desarrollo Nuclear (CEA\-DEN), Cubaenerg\'{\i}a, Cuba, and IAEA (International Atomic Energy Agency);
Swedish Research Council (VR) and Knut $\&$ Alice Wallenberg
Foundation (KAW);
Ukraine Ministry of Education and Science;
United Kingdom Science and Technology Facilities Council (STFC);
The United States Department of Energy, the United States National
Science Foundation, the State of Texas, and the State of Ohio;
Ministry of Science, Education and Sports of Croatia and  Unity through Knowledge Fund, Croatia;
Council of Scientific and Industrial Research (CSIR), New Delhi, India;
Pontificia Universidad Cat\'{o}lica del Per\'{u}.
\end{acknowledgement}

\bibliographystyle{utphys}
\bibliography{draft}{}

\newpage
\appendix
\section{The ALICE Collaboration}
\label{app:collab}

\begingroup
\small
\begin{flushleft}
J.~Adam$^{\rm 40}$, 
D.~Adamov\'{a}$^{\rm 84}$, 
M.M.~Aggarwal$^{\rm 88}$, 
G.~Aglieri Rinella$^{\rm 36}$, 
M.~Agnello$^{\rm 110}$, 
N.~Agrawal$^{\rm 48}$, 
Z.~Ahammed$^{\rm 132}$, 
S.U.~Ahn$^{\rm 68}$, 
S.~Aiola$^{\rm 136}$, 
A.~Akindinov$^{\rm 58}$
% ...

\bigskip 

\textbf{\Large Affiliation Notes} 

% ...
$^{\rm IV}$ Also at: M.V. Lomonosov Moscow State University, D.V. Skobeltsyn Institute of Nuclear, Physics, Moscow, Russia

\bigskip

\textbf{\Large Collaboration Institutes} 

% ...
$^{36}$ European Organization for Nuclear Research (CERN), Geneva, Switzerland\\
% ...
$^{40}$ Faculty of Nuclear Sciences and Physical Engineering, Czech Technical University in Prague, Prague, Czech Republic\\
% ...
$^{48}$ Indian Institute of Technology Bombay (IIT), Mumbai, India\\
% ...
$^{58}$ Institute for Theoretical and Experimental Physics, Moscow, Russia\\
% ...
$^{68}$ Korea Institute of Science and Technology Information, Daejeon, South Korea\\
% ...
$^{84}$ Nuclear Physics Institute, Academy of Sciences of the Czech Republic, \v{R}e\v{z} u Prahy, Czech Republic\\
% ...
$^{88}$ Physics Department, Panjab University, Chandigarh, India\\
% ...
$^{110}$ Sezione INFN, Turin, Italy\\
% ...
$^{132}$ Variable Energy Cyclotron Centre, Kolkata, India\\
% ...
$^{136}$ Yale University, New Haven, Connecticut, United States\\
% ...
\endgroup

\end{document}